\documentclass[twocolumn,aps,prl,amsmath,amssymb]{revtex4}
\usepackage{graphicx}% Include figure files
\usepackage{dcolumn}% Align table columns on decimal point
\usepackage{bm}% bold math
\usepackage[dvips]{color}

\begin{document}

\title{Graphene Q-switched, tunable fiber laser}

\author{D. Popa}
\author{Z. Sun}
\author{T. Hasan}
\author{F. Torrisi}
\author{F. Wang}
\author{A. C. Ferrari}
\email{acf26@eng.cam.ac.uk}
\affiliation{Department of Engineering, University of Cambridge,Cambridge CB3 0FA, UK}

\begin{abstract}
We demonstrate a wideband-tunable Q-switched fiber laser exploiting a graphene saturable absorber. We get$\sim$2$\mu$s pulses, tunable between 1522 and 1555nm with up to$\sim$40nJ energy. This is a simple and low-cost light source for metrology, environmental sensing and biomedical diagnostics.

\end{abstract}

\maketitle
Q-switching and mode-locking are the two main techniques enabling pulsed lasers\cite{Svelto1998}. In mode-locking, the random phase relation originating from the interference of cavity modes is fixed, resulting in a single pulse\cite{Svelto1998}, with typical duration ranging from tens ps to sub-10 fs\cite{Keller1996}, and a repetition rate corresponding to the inverse of the cavity round-trip time\cite{Keller1996}. In mode-locking, many aspects, including the dispersive and nonlinear proprieties of the intracavity components, need to be precisely balanced in order to achieve stable operation\cite{Svelto1998,Keller1996}. Q-switching is a modulation of the quality factor, Q, of a laser cavity\cite{Svelto1998}, Q being the ratio between the energy stored in the active medium and that lost per oscillation cycle\cite{Svelto1998} (thus, the lower the losses, the higher Q). In Q-switching, the active medium is pumped while lasing is initially prevented by a low Q factor\cite{Svelto1998}. The stored energy is then released in a pulse with duration ranging from $\mu$s to ns when lasing is allowed by a high Q factor\cite{Svelto1998}. The time needed to replenish the extracted energy between two consecutive pulses is related to the lifetime of the gain medium, which is typically$\sim$ms for erbium-doped fibres\cite{Svelto1998}. Thus the repetition rate of Q-switched lasers is usually low ($\sim$kHz\cite{Svelto1998}), much smaller than mode-locked lasers\cite{Svelto1998,Keller1996}. On the other hand, Q-switching enables much higher pulse energies and durations than mode-locking\cite{Svelto1998}. Q-switching has advantages in terms of cost, efficient operation (i.e. input power/output pulse energy) and easy implementation, compared to mode-locking, which needs a careful design of the cavity parameters to achieve a balance of dispersion and nonlinearity\cite{Svelto1998,Keller1996}. Q-switched lasers are ideal for applications where ultrafast pulses ($<$1ns) are not necessary, or long pulses are advantageous\cite{Paschotta1999,Siniaeva2009}, such as material processing, environmental sensing, range finding, medicine and long-pulse nonlinear experiments\cite{Paschotta1999,Laroche2002,Siniaeva2009}.

Q-switching can be active (exploiting, e.g., an acousto-optic or electro-optic modulator\cite{Svelto1998}), or passive (using, e.g., a saturable absorber (SA)\cite{Svelto1998}). Passive Q-switching features a more compact geometry and simpler setup compared to active, which requires additional switching electronics\cite{Svelto1998}. For Q-switching the SA recovery time does not need to be shorter than the cavity round-trip time, since the pulse duration mainly depends on the time needed to deplete the gain after the SA saturates\cite{Svelto1998,Keller1996}, unlike mode-locking\cite{Keller1996}. Doped bulk crystals\cite{Laroche2002}, and semiconductor saturable absorber mirrors (SESAMs)\cite{Kivisto2008,Paschotta1999} are the most common SAs in passive Q-switching\cite{Svelto1998}. However, the use of doped crystals as SAs requires extra elements (mirrors, lenses) to focus the fiber output into the crystal\cite{Laroche2002}. SESAMs have limited operation bandwidth, typically few tens nm\cite{Okhotnikov2004}, thus are not suitable for broad-band tunable pulse generation. Broadband SAs enabling easy integration into an optical fiber system are thus needed to create a compact Q-switched fibre laser.
\begin{figure}
\centerline{\includegraphics[width=70mm]{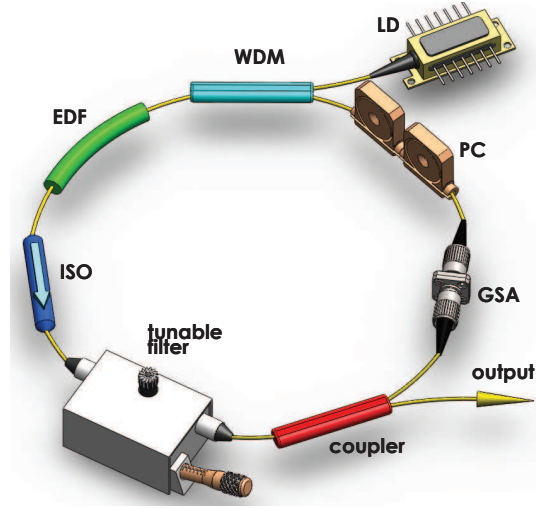}}
\caption{Setup: laser diode (LD), wavelength division multiplexer (WDM), erbium-doped fiber (EDF), isolator (ISO), graphene SA (GSA), polarization controller (PC)}
\label{setup}
\end{figure}

Single wall carbon nanotubes (SWNTs) and graphene are ideal SAs, due to their low saturation intensity, low cost and easy fabrication\cite{Wang2008,Sun2010g,Hasan2009,Set2004,Schibli2005,Rozhin2006pss,Martinez_oe_10,Solodyankin2008,Sun2008,Sun2009hp,Sun2010tg,sunnanores_10,Bonaccorso2010,Popa2010,Hasan2010}. Broadband operation is achieved in SWNT using a distribution of tube diameters\cite{Wang2008,Hasan2009}, while this is an intrinsic property of graphene, due to the gapless linear dispersion of Dirac electrons\cite{Geim2007,Bonaccorso2010,Sun2010g,Sun2010tg,Hasan2010}. Q-Switching was reported using SWNTs: Ref.\onlinecite{Da-Peng2010} achieved 14.1nJ pulse energy and 7$\mu$s width, while Ref.\onlinecite{Dong2010} 13.3nJ and 700ns. After the demonstration of a graphene-based mode-locked laser\cite{Hasan2009}, various group implemented graphene SA in a variety of mode-locked cavity designs\cite{Popa2010,Bonaccorso2010,Hasan2010,Sun2010g,Sun2010tg,Zhang2009,Song2010,Martinez2010}.

Here, we demonstrate a fiber laser Q-switched by a graphene saturable absorber (GSA). The broadband absorption of graphene enables Q-switching over a 32nm range, limited only by our tunable filter, not graphene itself. The pulse energy is$\sim$40nJ, for$\sim$2$\mu$s duration.

\begin{figure}
\centerline{\includegraphics[width=75mm]{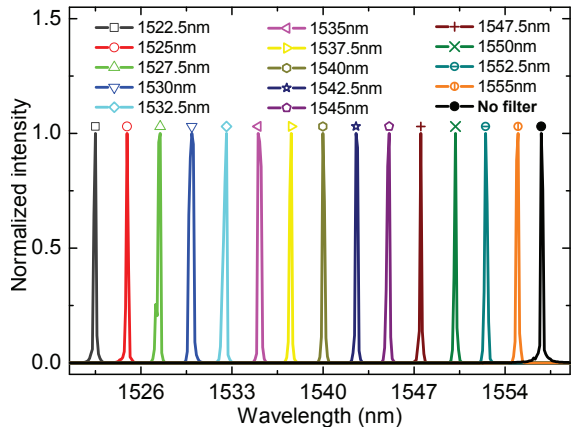}}
\caption{Output spectra for 14 tuning wavelengths. The curve with a filled circle corresponds to Q-switching without filter.}
\label{spectra}
\end{figure}
Graphite flakes are exfoliated by mild ultrasonication with sodium deoxycholate (SDC)\cite{Sun2010g,Hasan2010,Hernandez2008}. A dispersion enriched with single (SLG) and few layer graphene (FLG)\cite{Hasan2010} is then mixed with an aqueous solution of polyvinyl alcohol (PVA). After water evaporation, a$\sim$50$\mu$m thick graphene-PVA composite is obtained\cite{Hasan2009,Sun2010g}. This is then placed between two fiber connectors to form a fiber-compatible SA, then integrated into a laser cavity, Fig.\ref{setup}, with 1.25m erbium doped fiber (EDF) as gain medium, pumped with a 980nm laser diode (LD), coupled via a wavelength division multiplexer (WDM). An optical isolator (ISO) ensures unidirectional light propagation. An in-line tunable optical bandpass filter is inserted after the ISO. Our EDF can support lasing between 1520 and 1560nm\cite{Agrawal2001}. The operation wavelength is selected rotating the dielectric interference filter. The 20\% port of an optical coupler provides the laser output. The rest of the cavity consists of a combination of single mode fiber (SMF) Flexcor 1060 and SMF-28. All fibers used in our cavity are polarization-independent, i.e. they support any light polarization, even if this changes as a result of outside perturbations (e.g. mechanical stresses, bending, or temperature). Thus, to improve the output pulse stability, we place in the cavity a polarization controller (PC), consisting of 2 spools of SMF-28 fiber acting as retarders. The total retardation induced by the PC is a function of the fiber geometry in the spool\cite{Agrawal2001}. This allows to maintain a polarization state after each round trip. The total cavity length is$\sim$10.4m. The operation is evaluated by a 14GHz bandwidth photo-detector and an oscilloscope. A spectrum analyzer with 0.07nm resolution measures the output spectrum.
\begin{figure}
\centerline{\includegraphics[width=75mm]{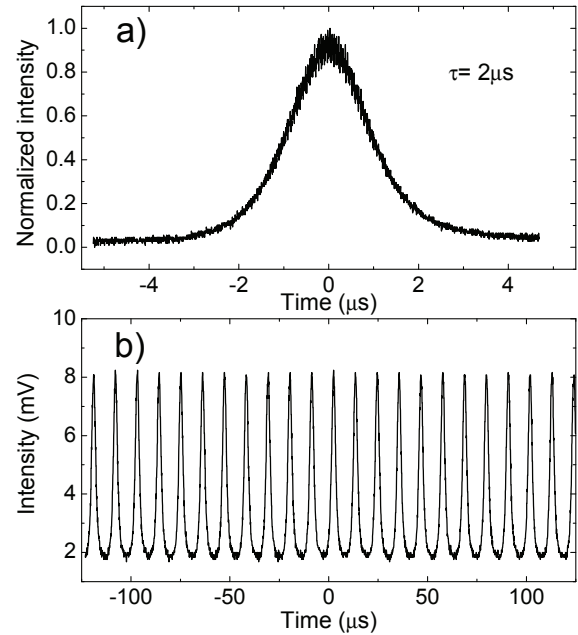}}
\caption{a) Single pulse envelope. b) Typical pulse train for 2.8mW output power.}
\label{pulse}
\end{figure}

\begin{figure}
\centerline{\includegraphics[width=75mm]{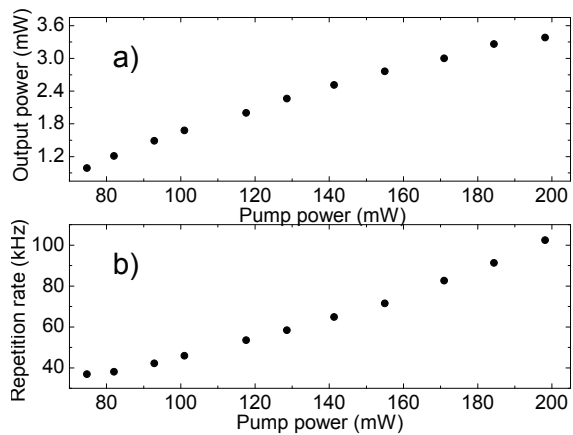}}
\caption{(a)Output power and (b) repetition rate,  as a function of input pump power at 1540nm}
\label{rr}
\end{figure}

Continuous wave (CW) operation starts at$\sim$43mW pump power; pulsed operation at$\sim$74mW. The repetition rate is pump-dependent up to$\sim$200mW (Fig.\ref{rr}b), a typical signature of Q-switching\cite{Svelto1998}. The output spectrum is tunable from$\sim$1522 to 1555nm. This is comparable to the 31nm range reported for doped crystal Q-switched tunable lasers\cite{Laroche2002}, but much larger than the 5nm thus far achieved for SWNT Q-switched lasers\cite{Da-Peng2010,Dong2010}. Our tuning range is limited by the filter and by the EDF gain, not the GSA. Fig. \ref{spectra} shows the output spectra for 14 wavelengths at$\sim$2.5mW output power.  Without filter, the laser exhibits Q-switching at 1557nm. The full width at half maximum (FWHM) spectral width is 0.3$\pm$0.1nm over the whole tuning range, much shorter than thus far achieved for graphene mode-locked lasers\cite{Bonaccorso2010,Popa2010,Hasan2010,Sun2010g,Sun2010tg,Zhang2009,Song2010,Martinez2010}.

Fig.\ref{pulse}a plots a typical pulse envelope, having FWHM$\sim$2$\mu$s, comparable to fiber lasers Q-switched with other SAs (e.g. SESAMs\cite{Kivisto2008,Paschotta1999}, doped crystals\cite{Laroche2002}, and SWNTs\cite{Da-Peng2010,Dong2010}), but much longer than thus far achieved in graphene mode-locked fiber lasers\cite{Bonaccorso2010,Popa2010,Hasan2010,Sun2010g,Sun2010tg,Zhang2009,Song2010,Martinez2010}. The output pulse duration has little dependence on wavelength, possibly due to the flat gain coefficient of our EDF\cite{Agrawal2001}. Fig.\ref{pulse}b shows the pulse train for a typical laser output at 158mW pump power.

The output power varies from 1 to 3.4mW as a function of pump power. The slope efficiency, i.e. the slope of the line obtained by plotting the laser output power against the input pump power\cite{Svelto1998}, is$\sim2\%$. The repetition rate as a function of pump power varies from 36 to 103KHz (Fig.\ref{rr}b), with a 67KHz change for a 2.4mW output power variation. Unlike mode-locked lasers, where the repetition rate is fixed by the cavity length\cite{Svelto1998}, in Q-switched lasers this depends on pump power\cite{Svelto1998}. As this increases, more gain is provided to saturate the SA and, since pulse generation relies on saturation, the repetition rate increases with pump power\cite{Svelto1998}. The maximum output pulse energy is$\sim$40nJ for$\sim$60KHz repetition rate, similar to that achieved using other SAs\cite{Dong2010}. Compared to graphene mode-locked fiber lasers\cite{Bonaccorso2010,Popa2010,Hasan2010,Sun2010g,Sun2010tg,Zhang2009,Song2010,Martinez2010}, our pulse energy is$\sim$6 times larger, but with less peak power, due to the larger pulse duration. It is also much larger than thus far achieved in SWNT Q-switched lasers\cite{Da-Peng2010,Dong2010}. Even higher energies, thus peak powers, could be enabled by evanescent field interaction with GSA\cite{Song2010} and high-gain fibers (e.g. cladding-pumped fibers\cite{Laroche2002} or large mode area fibers\cite{Paschotta1999}).

\begin{figure}
\centerline{\includegraphics[width=75mm]{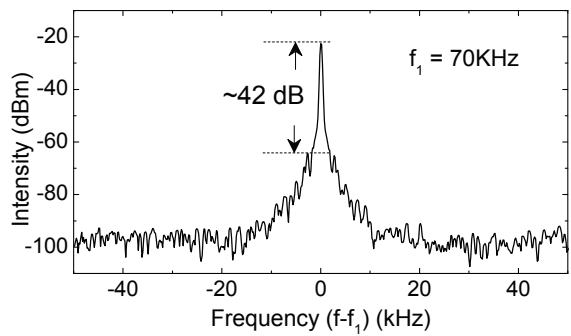}}
\caption{RF spectrum measured around$\sim$70KHz at 1540nm}
\label{rf}
\end{figure}

The radio-frequency (RF) measurement of the output intensity at 70Hz, corresponding to a period of$\sim$143$\mu$s, is shown in Fig.\ref{rf}. The peak to pedestal extinction is$\sim$40dB (10$^4$ contrast), confirming pulse stability.

In conclusion, we achieved Q-switching exploiting a graphene-based saturable absorber, using standard, telecom grade, fibre components. The wideband operation of graphene enables broad band tunability. Such wideband Q-switched laser could provide a simple, low-cost, and convenient light source for metrology, environmental sensing and biomedical diagnostics.

We acknowledge funding from EPSRC GR/S97613/01, EP/E500935/1, ERC NANOPOTS, a Royal Society Brian Mercer Award for Innovation, King's College.

\end{document}